\documentclass[prb,aps,twocolumn,superscriptaddress,showpacs]{revtex4}

\usepackage{graphicx}

\begin{document}

\title{Magnetic field resonantly enhanced free spins in heavily underdoped YBa$_{2}$Cu$_{3}$O$_{6+x}$}

\author{C. Stock}
\affiliation{ISIS Facility, Rutherford Appleton Labs, Chilton, Didcot, OX11 0QX, UK}

\author{W. J. L. Buyers}
\affiliation{National Research Council, Chalk River, Ontario, Canada K0J 1JO}
\affiliation{Canadian Institute of Advanced Research, Toronto, Ontario, Canada M5G 1Z8}

\author{K. C. Rule}
\affiliation{Helmholtz Zentrum Berlin, D-14109, Berlin, Germany}

\author{J.-H. Chung}
\affiliation{NIST Center for Neutron Research, Gaithersburg, Maryland USA 20899}
\affiliation{Department of Phyics, Korea University, Seoul, 136-701, Korea}

\author{R. Liang}
\affiliation{Physics Department, University of British Columbia, Vancouver, B. C., Canada V6T 2E7}
\affiliation{Canadian Institute of Advanced Research, Toronto, Ontario, Canada M5G 1Z8}

\author{D. Bonn}
\affiliation{Physics Department, University of British Columbia, Vancouver, B. C., Canada V6T 2E7}
\affiliation{Canadian Institute of Advanced Research, Toronto, Ontario, Canada M5G 1Z8}

\author{W. N. Hardy}
\affiliation{Physics Department, University of British Columbia, Vancouver, B. C., Canada V6T 2E7}
\affiliation{Canadian Institute of Advanced Research, Toronto, Ontario, Canada M5G 1Z8}

\date{\today}

\begin{abstract}

Using neutron scattering, we investigate the effect of a magnetic field on the static and dynamic spin response in heavily underdoped superconducting YBa$_{2}$Cu$_{3}$O$_{6+x}$ (YBCO$_{6+x}$) with x=0.33 (T$_{c}$=8 K) and 0.35 (T$_{c}$=18 K).  In contrast to the heavily doped and superconducting monolayer cuprates, the elastic central peak characterizing static spin correlations does not respond observably to a magnetic field which suppresses superconductivity.  Instead, we find a magnetic field induced resonant enhancement of the spin fluctuations.  The energy scale of the enhanced fluctuations matches the Zeeman energy within both the normal and vortex phases while the momentum dependence is the same as the zero field bilayer response.  The magnitude of the enhancement is very similar in both phases with a fractional intensity change of $(I/I_{0}-1) \sim 0.1$.  We suggest that the enhancement is not directly correlated with superconductivity but is the result of almost free spins located near hole rich regions.

\end{abstract}

\pacs{74.72.-h, 75.25.+z, 75.40.Gb}

\maketitle

\section{Introduction}

The cuprate superconductors are all based on doping charge into the magnetic CuO$_{2}$ planes of a Mott insulator.~\cite{Kastner98:70,Birgeneau06:75,Buyers06:385}  A direct interplay between the magnetism of the Cu$^{2+}$ spins and the electronic response occurs for hole doped cuprates where long-range antiferromagnetism is suppressed in favor of superconductivity at a critical hole concentration of p$_{c}$=0.055.~\cite{Fujita02:65, Wakimoto01:63}  Whether antiferromagnetic fluctuations can account for the pairing mechanism in the cuprates is still a matter of debate although many studies have shown that antiferromagnetism and superconductivity are indeed coupled.~\cite{Wakimoto07:98,Wakimoto04:92}   

The application of a magnetic field allows the superconducting order parameter to be suppressed continuously without change in the chemical composition.  Much work has been done on the monolayer La$_{2-x}$Sr$_{x}$CuO$_{4}$ (LSCO) system where the effects of a magnetic field vary with hole doping.  It is by no means obvious, based on the available data, that the entire hole doping phase diagram is described by a common response with a universal physical origin.  For lightly doped, non superconducting and insulating concentrations, a magnetic field was found to suppress the elastic scattering and was associated with a reorientation of the Cu$^{2+}$ spin direction.~\cite{Matsuda02:66}  Within the superconducting phase of nearly optimally doped LSCO and oxygen ordered La$_{2}$CuO$_{4+y}$ an enhancement occurs in the static long-ranged antiferromagnetism on application of a magnetic field.~\cite{Katano00:62,Khaykovich02:66,Khaykovich03:67,Lake05:4}  Few results have been reported for less than optimally doped compositions which are superconducting, although  one study on La$_{2-x}$Ba$_{x}$CuO$_{4}$ (LBCO) x=0.095 showed no change in the elastic scattering in a magnetic field.~\cite{Dunsiger08:77}  Initially, the increase in magnetic order for heavily doped samples was suggested to be the result of antiferromagnetism within the vortex cores (Ref. \onlinecite{Lake02:415}). It has since been argued that the results are better described in terms of the close proximity of a quantum critical point separating a purely superconducting phase from a phase where superconductivity and spin density wave order coexist.~\cite{Demler01:87,Zhang02:66}  This model predicts that there is an upper hole concentration where a threshold field is required before an enhancement of antiferromagnetism occurs as has been observed experimentally.~\cite{Khaykovich05:71}  The doping and magnetic field dependence summarized here has been reconciled by a combined muon and neutron study of LSCO near a hole doping of $x \sim 1/8$.~\cite{Chang08:78}  The enhancement of static antiferromagnetism was measured to be most pronounced near hole concentrations of $1/8$ with the enhancement decreasing with lower doping.  It was concluded that the effect of the field is to drive the system toward the $1/8$ ground state.

In comparison to the monolayer cuprates, relatively few magnetic field experiments have been reported on the bilayer YBCO$_{6+x}$ system.  On suppressing superconductivity with a magnetic field in YBCO$_{6.6}$, a concomitant decrease in intensity of the inelastic resonance peak was observed with the field rotated $\sim$ 21$^{\circ}$ from the [001] axis.~\cite{Dai00:406}  The possibility of ordered magnetism within the vortex cores has been investigated in optimally doped YBCO$_{7-\delta}$ (T$_{c}$=90 K) with the field aligned along the [1$\overline{1}$0] direction and some evidence was claimed for weak ferromagnetic ordering.~\cite{Vaknin00:329} There has been little work on the effects of a magnetic field on the excitation spectrum or the static magnetism in hole doped cuprates beyond the monolayer system described above.

We have carried out neutron inelastic scattering studies of the entire spin excitation spectrum of heavily underdoped superconducting YBCO$_{6+x}$.~\cite{Stock06:73,Stock07:75,Stock08:77,Yamani07:460} While the heavily doped YBCO$_{6+x}$ exhibits a well defined resonance peak and low-energy incommensurate scattering~\cite{Fong00:61,Dai01:63,Stock04:69}, the inelastic spectrum in heavily underdoped and superconducting YBCO$_{6+x}$ is very different and several studies have been devoted to intermediate oxygen concentrations.~\cite{Li08:77,Hinkov08:319}  In YBCO$_{6.35}$ (T$_{c}$=18 K) and YBCO$_{6.33}$ (T$_{c}$=8 K), the low-energy magnetic response consists of a central peak and a broad inelastic feature peaked at $\hbar \omega$ $\sim$ 2 meV.   As established in YBCO$_{6.35}$,~\cite{Stock08:77} the central peak sets in over a broad temperature range while the low-energy spectral weight is suppressed over a similar temperature range suggesting that spectral weight is conserved with the central peak intensity being removed from low-energy spin fluctuations.

We now describe how the spin fluctuations in heavily underdoped superconducting YBCO$_{6+x}$ respond to magnetic fields that suppress the superconducting order parameter.  We find no observable response of the elastic central peak to a magnetic field both within the vortex and normal states.  Instead we demonstrate that a resonant enhancement of the low-energy spin fluctuations takes place at an energy scale similar to the Zeeman energy.  We speculate that this effect is due to the magnetization of weakly coupled spins  located near hole rich regions where exchange fields are small.

\section{Experiment}

\subsection{Sample and instrument details}

Measurements were made on two systems.  The first sample studied on SPINS (National Institute of Standards and Technology, NIST) consisted of seven $\sim$ 1 cc crystals of YBCO$_{6.35}$ (T$_{c}$=18 K) coaligned in the (HHL) scattering plane.  A second set of measurements was conducted at FLEX (Helmholtz Zentrum Berlin, HZB) with four $\sim$ 1 cc crystals of YBCO$_{6.33}$ (T$_{c}$=8 K) also coaligned in the (HHL) scattering plane.  The YBCO$_{6.33}$ (YBCO$_{6.35}$) materials were orthorhombic with lattice constants $a$=3.844 (3.843) \AA, $b$=3.870 (3.871) \AA \ and $c$=11.791 (11.788) \AA \ from which a hole doping of $p$=0.055 (0.060) was derived based on the lattice constants and superconducting transition temperatures.~\cite{Liang06:73,Tallon95:51}    

The magnetic field directions on SPINS and FLEX cold triple axis spectrometers were vertical and horizontal respectively.  The SPINS results were for YBCO$_{6.35}$ (T$_{c}$=18 K) in a 11 T vertical field with the field aligned along the [1$\overline{1}$0] axis.  In the FLEX experiment a 6 T horizontal magnetic was aligned along the [001] axis of YBCO$_{6.33}$ (T$_{c}$=8 K).  In both experiments the monochromators were vertically focussed PG(002) crystals. SPINS data were collected with a final energy of E$_{f}$=3.7 meV and a  beryllium oxide filter was placed after the sample in parallel with a radial collimator.  The analyzer was horizontally focussed PG(002) with a 5$^{\circ}$ acceptance. On FLEX, the final energy was E$_{f}$=2.9 meV with a beryllium filter placed before the sample.  The PG(002) analyzer was horizontally focussed  with a 3$^{\circ}$ acceptance for the inelastic scattering and was flat with a collimation sequence of guide-60$'$-$S$-60$'$-open for elastic scattering.  The energy resolutions defined as the full-width at half maximum at the elastic line were $\delta E$=0.08 and 0.14 meV for the FLEX and SPINS experiments respectively.

\begin{figure}[t]
\includegraphics[width=75mm]{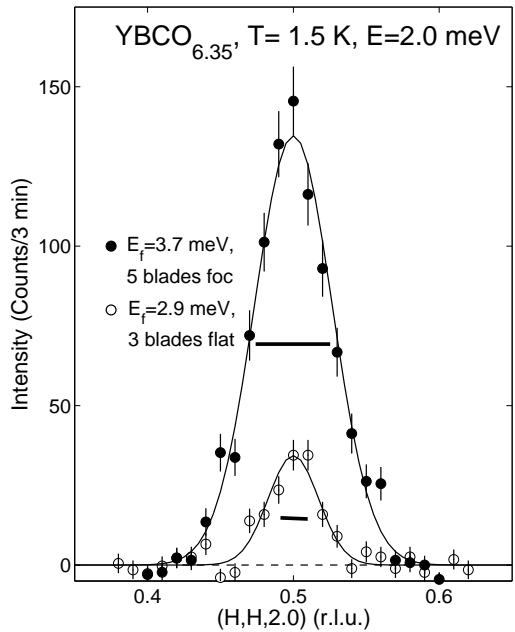}
\caption{Constant E=2.0 meV scans at $\vec{Q}$=(1/2,1/2,2) with two different experimental configurations on the SPINS cold triple-axis spectrometer.  The open circles were obtained with E$_{f}$=2.9 meV and a flat analyzer.  The filled circles illustrate the same scan with E$_{f}$=3.7 meV and a horizontally focussed analyzer with 5$^{\circ}$ acceptance.  The latter configuration gave a large gain in intensity (with a loss in momentum resolution) and was used for the magnetic field experiment on SPINS.} \label{exp_config}
\end{figure}

Initial zero field data on YBCO$_{6.35}$ were taken on SPINS with a flat analyzer and E$_{f}$=2.9 meV.  This configuration worked well for characterizing the low-energy spin fluctuations at zero field, but we found a large gain in intensity by selecting a configuration with E$_{f}$=3.7 meV and a horizontally focussed analyzer.  A comparison (Fig. \ref{exp_config}) normalized by counting time illustrates a significant gain in the integrated intensity at the cost of coarsening the momentum resolution (denoted by the horizontal bars).  The horizontally focussed analyzer configuration was chosen both at SPINS and FLEX to determine the magnetic field dependence of the low-energy Cu$^{2+}$ spin fluctuations. 

Because of the Meissner effect, superconductors experience large forces when placed in a changing magnetic field. In all experiments the sample was heated to 30 K (well above the onset of superconductivity in both concentrations) before any change in field.  To check whether the sample had moved, the filters were removed and the (1,1,4) Bragg peak was scanned using $\lambda/2$ neutrons at the same spectrometer angles where the magnetic scattering was measured.  While no change was observed in the nuclear Bragg peaks in the SPINS experiment, temperature independent changes of up to 10$\%$ of the Bragg peak intensity were observed in the FLEX horizontal field experiment.  This change in intensity followed the same trend with field as did the elastic scattering.  Because of the larger error bars, this effect was not noticeable for the inelastic scans.  The conclusions were drawn from elastic scans with the horizontal magnetic field aligned both along the [001] and [00$\overline{1}$] axes.  They showed opposite effects on the elastic intensity of the (114) nuclear peak and the (1/2 1/2 2) elastic magnetic peak of equal proportion.   We have corrected the elastic magnetic scattering (Fig. \ref{YBCO_elastic}) for the nuclear Bragg peak intensity in a field.  As an extra precaution, all \textit{inelastic} data from FLEX are an average between the field aligned along the [001] and the [00$\overline{1}$] axes.

\subsection{Magnetization in cuprate superconductors}

While the field orientation and transition temperatures (and hence critical fields) in the experiments are different, the actual magnetization along the [001] direction is likely to be very similar.  This can be seen from torque magnetometry in  YBCO$_{6+x}$ and Bi$_{2}$Sr$_{2}$CaCu$_{2}$O$_{8+x}$ where even a small component of the field along the [001] axis can result in a significant magnetization along $c$.~\cite{Farrell89:63,Farrell90:64,Martinez92:69,Tuominem90:42}  This arises because screening currents prefer to form in the $a-b$ CuO$_{2}$ planes rather than with a component along $c$.  In the regime $H_{c1}<<H<<H_{c2}$, the ratio of the transverse ($M_{T}$, perpendicular to the field direction) to longitudinal ($M_{L}$, parallel to the field direction) magnetization has been found to be given by

\begin{eqnarray}
{M_{T} \over M_{L}}=(\gamma-1) {{\sin(\theta) \cos(\theta)} \over {\sin^{2}(\theta)+\gamma \cos^{2}(\theta)}},
\end{eqnarray}

\noindent where $\gamma$ is the effective mass ratio $m_{L}/m_{T}$ and $\theta$ is the angle of the applied magnetic field with respect to the [001] axis.~\cite{Tuominem90:42}  Experiments on optimally doped YBCO$_{7-\delta}$ (T$_{c}$=90 K) give $\gamma$ $\sim$ 30 $\pm$ 5.~\cite{Farrell90:64}  Experiments on Bi$_{2}$Sr$_{2}$CaCu$_{2}$O$_{y}$ gave a much larger value of 280 $\pm$ 20.~\cite{Farrell89:63,Martinez92:69}  The most relevant experimental work on the magnetic penetration depth in heavily underdoped YBCO$_{6+x}$ shows that the anisotropy increases with decreasing doping and reaches $\gamma \sim \lambda_{c}/\lambda_{ab} \sim 100$ in the region of the present experiments.~\cite{Hosseini04:03,Broun07:99,Liang05:94}

For the [1$\overline{1}$0] field experiment on YBCO$_{6.35}$, the crystal axes and field direction were only oriented within $\pm$ 1.5$^{\circ}$ being determined largely by the mosaic spread of the samples.  Substituting $\gamma$=100 and $\theta$=88.5$^{\circ}$ in the above equation gives $M_{T}$/M$_{L}$ $\sim$ 2.  Therefore, even though the field is nominally aligned within the $a-b$ plane (along [1$\overline{1}$0]), a significant magnetization along the [001] direction results from the large anisotropy $\gamma$.  With a field of 11 T, the magnetization along [001] is comparable with the 6 T field oriented along the [001] direction.

The upper critical field $H_{c2}$ characterizes how much the superconducting order parameter is suppressed.  $H_{c2}$ has been studied as a function of hole doping in underdoped YBCO$_{6+x}$ by Gantmakher \textit{et al.} using resistivity and with magnetic field aligned along the [001] axis.~\cite{Gantmakher99:88}  Based on the resistivity data, we estimate that H$_{c2}$ for YBCO$_{6.35}$ (T$_{c}$=18 K) is at least 15 T at T=2 K, for the magnetic field aligned [001].  For YBCO$_{6.33}$ (T$_{c}$=8 K), the upper critical field is estimated to be 4 $\pm$ 0.5 T (at 1.5 K) for the field applied along the $c$ axis.   Therefore, for the horizontal field experiment conducted on FLEX with 6 T applied along the [001] axis of YBCO$_{6.33}$ the samples were in the normal state.  The SPINS experiment on YBCO$_{6.35}$ with a vertical field of 11 T, however, the material was in the vortex state.  

\section{YBCO$_{6.35}$, T$_{c}$=18 K, $\mu_{0}H$=11 T $\parallel$ [1$\overline{1}$0] - vortex state}

\begin{figure}[t]
\includegraphics[width=75mm]{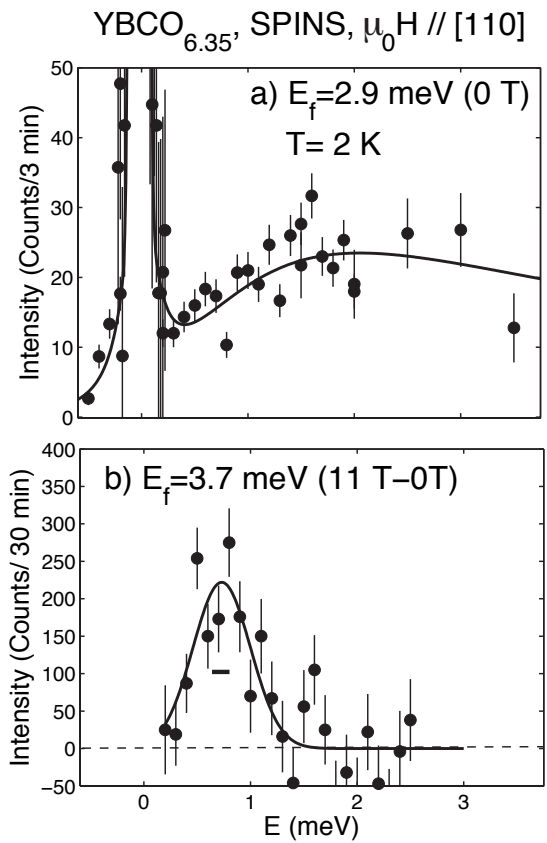}
\caption{Constant $\vec{Q}$=(1/2,1/2,2.0) scans conducted using the SPINS cold triple-axis spectrometer.  The sample was YBCO$_{6.35}$ (T$_{c}$=18 K).   The form of the low-energy zero field spectrum is illustrated in panel $a)$ at 2 K with a flat analyser and E$_{f}$=2.9 meV.  A subtraction of 0 T data from 11 T scans is illustrated in panel $b)$ with a vertical field aligned along the [1$\overline{1}$0] direction.} \label{YBCO_635_energy}
\end{figure}

\begin{figure}[t]
\includegraphics[width=75mm]{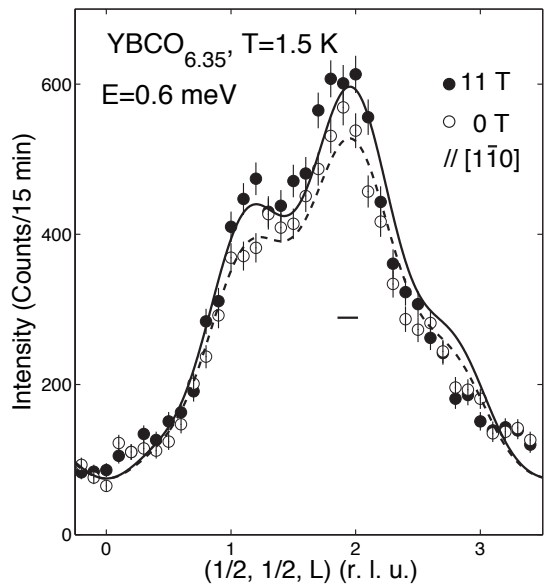}
\caption{Constant E=0.6 meV scans at $\mu_{0}H$=0 and 11 T with field parallel to [1$\overline{1}$0] direction.  The sample was YBCO$_{6.35}$ (T$_{c}$=18 K).  The lines through the data are described in the text.  The data was obtained on SPINS using a horizontally focussed analyzer and E$_{f}$=3.7 meV.} \label{L_scan}
\end{figure}

The low-energy spin response of YBCO$_{6.35}$ to an 11 T field aligned near the [1$\overline{1}$0] direction is illustrated in Fig. \ref{YBCO_635_energy}. The previously published zero field results are displayed in panel $a)$ and the field subtracted data determined at lower resolution are plotted in panel $b)$.  On application of a field, a resonant enhancement at $\sim$ 0.6 meV is observed at low-energies.   The relative increase at E=0.6 meV at 11 T was measured to be $I(H)/I(0)-1 \sim 0.1$ (derived from scans in momentum similar to those displayed in Fig. \ref{L_scan} and Fig. \ref{exp_config}).   The field induced resonance is not resolution limited (as indicated by the horizontal resolution bar) and indicates strong damping.

To investigate how the intensity changes as a function of momentum transfer, we conducted constant E=0.6 meV scans along the [001] direction.  The scan along (1/2,1/2,L) is sensitive to the sign of the spin correlations along the $c$-axis.  Magnetic scattering indicative of ferromagnetic spin correlations between nearest neighbor planes would result in scattering at L=0, whereas a non-zero cross section at L $\sim$ 1.7 is characteristic of antiferromagnetic correlations as shown in Fig. \ref{L_scan}. The solid and dashed lines through the data are fits to the following,

\begin{eqnarray}
{I(L)}={C+A\sin^2(\pi L(1-2z))(1+B \cos(2\pi L)))}.
\end{eqnarray} 

\noindent Here $A$ is an amplitude, $(1-2 z)$ represents the bilayer spacing with $z$=0.36, $B$ represents a measure of the correlations between neighboring bilayers, and $C$ is an overall constant representative of the nonmagnetic background. The scan along L illustrates that the broad enhancement near L $\sim$ 1.7 follows the bilayer structure factor and therefore corresponds to antiferromagnetically correlated spins between neighboring CuO$_{2}$ planes.  We find no measurable enhancement at L=0 indicating the absence of ferromagnetic correlations between neighboring CuO$_{2}$ layers.  This result contrasts with that previously reported in more heavily doped YBCO$_{6.5}$.~\cite{Vaknin00:329}  Because the enhancement itself follows the same pattern in L as the bilayer structure factor, it follows that the field induced spin changes lie within the bilayers and any role for spins in the chains is excluded.  

In our previous analysis and experiments at larger energy transfers (E $\sim$ 2.5 meV above the peak in the broad inelastic response)~\cite{Stock08:77} we did not require a term representing correlations between neighboring bilayers and found the data to be well described by the above expression with $B$=0.  Scans at energy transfers below the broad maximum at $\sim$ 2 meV in the inelastic response illustrate the presence of stronger bilayer correlations.  Despite the fractional increase of $\sim$ 10 $\%$ in the low-energy spectral weight on application of a magnetic field, we observed no significant change in the elastic intensity within a sensitivity of $I(H)/I(0)-1 \sim$ 0.05 at the lowest temperatures.  The elastic scattering from static spin correlations will be discussed in the following section on YBCO$_{6.33}$ (T$_{c}$=8 K) where the superconducting order parameter is completely suppressed with a horizontal field aligned along the [001] axis.

\section{YBCO$_{6.33}$, T$_{c}$=8 K, $\mu_{0}H$=6 T $\parallel$ [001] - normal state}

We now present results for YBCO$_{6.33}$ (T$_{c}$=8 K) in a horizontal field along the [001] axis.  This sample has a lower hole doping and a lower superconducting transition temperature than the sample discussed above.  Through having the field aligned along the $c$ axis and with the lower T$_{c}$ we are now able to suppress entirely the superconducting order parameter.  The measurements were taken on the FLEX cold triple-axis spectrometer with a 6 T field aligned along the [001] axis.  We first present data on the elastic scattering and then the dynamics probed through inelastic scattering. 

\subsection{Elastic Scattering}

\begin{figure}[t]
\includegraphics[width=85mm]{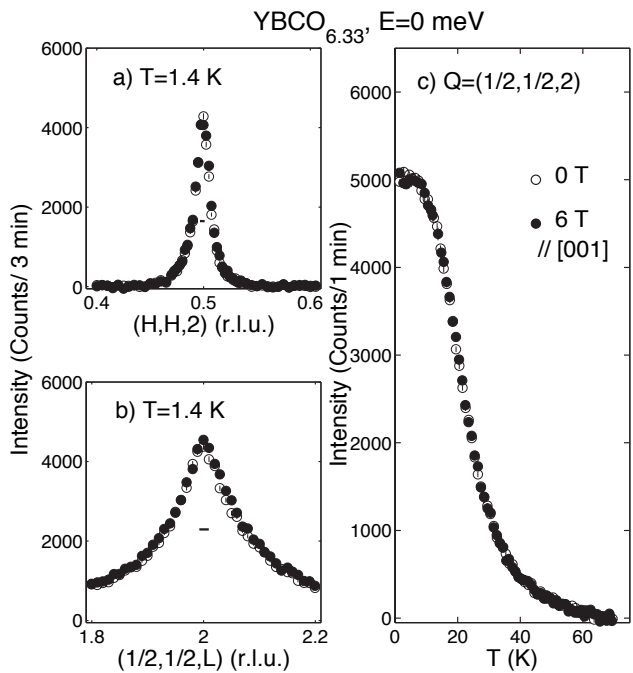}
\caption{Elastic scattering near $\vec{Q}$=(1/2,1/2,2.0) corrected for a nonmagnetic background taken at 80 K as measured with the FLEX
cold triple-axis spectrometer.  The temperature dependence in both 0 (superconducting state) and 6 T (normal state) fields is plotted in panel $c)$.  Panels $a)$ and $b)$ were taken with a flat analyzer and panel $c)$ with a horizontally focussed analyser.  The data at each field has been scaled to the (114) Bragg peak intensity measured with $\lambda/2$ neutrons in the absence of Be filters.  The error bars are the size of the data points.  There is no significant field effect on the static spin correlations.} \label{YBCO_elastic}
\end{figure}

The effect of a magnetic field on the elastic scattering, which arises from correlations that fluctuate on a timescale longer than $\tau$ $\sim$ $2\hbar/\delta E$ $\sim$ 30 ps, is shown in Fig. \ref{YBCO_elastic}.  Panels $a)$ and $b)$ illustrate the magnetic scattering at T=1.4 K with applied magnetic fields of $\mu_{0}H$=0 and 6 T applied along the $c$ axis.  A background measured at 80 K has been subtracted.  The peak temperature dependence at 0 and 6 T is shown in panel $c)$.  Within our experimental uncertainties (of order 2$\%$) we find no change in the temperature dependence, intensity, or lineshape of the static spin correlations.  Based on this analysis, we therefore conclude that in the heavily underdoped regime, suppression of the superconducting order parameter does not have a strong effect on the central peak over the field range studied.

This result contrasts strongly with the behavior of the monolayer LSCO system where, at least near optimal doping, a clear field enhancement is observed.~\cite{Khaykovich02:66,Lake05:4}  The relative enhancement, characterised by $I(H)/I(0)-1$, in all experiments in the monolayer cuprates range from 0.1 (for stage-6 La$_{2}$CuO$_{4+y}$) to 1.5 (in La$_{1.895}$Sr$_{0.105}$CuO$_{4}$) for fields up to 15 T applied along the $c$-axis.  The experiment presented here is certainly sensitive enough to detect changes of the order measured in the monolayer cuprates.  We note that H$_{c2}$ for those experiments was $\sim$ 30 T.

\subsection{Inelastic Scattering}

\begin{figure}[t]
\includegraphics[width=75mm]{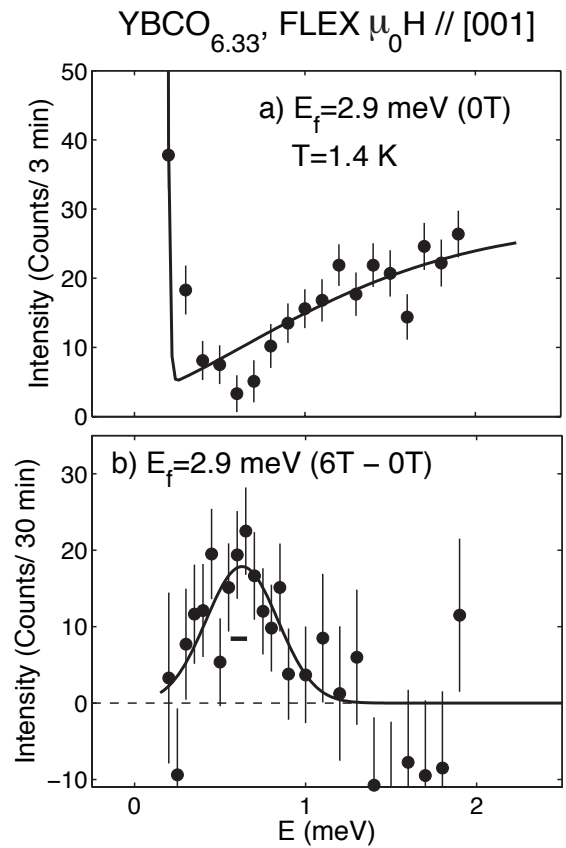}
\caption{Constant $\vec{Q}$=(1/2, 1/2, 2.0) scans with the FLEX cold triple-axis spectrometer on YBCO$_{6.33}$  (T$_{c}$=8 K).  The low-energy zero field spectrum is illustrated in panel $a)$ at 1.4 K.  A subtraction of 0 T data from 6 T scans is illustrated in panel $b)$.  The data was obtained using a horizontal magnet with the field applied along the [001] direction.} \label{YBCO_633_energy}
\end{figure}

The effect of the 6 T magnetic field along [001] on the low-energy fluctuations of the Cu$^{2+}$ is plotted in Fig. \ref{YBCO_633_energy}.  The zero field spectrum below 2 meV is shown in panel $a)$.  The data was obtained by fixing $Q$ and scanning energy transfer at $\vec{Q}$=(1/2,1/2,2.0) while using scans at $\vec{Q}$=(0.7,0.7,2.0) and (0.3,0.3,2.0) as the background.  This method of background subtraction is similar to that used in studying YBCO$_{6.35}$ (Refs. \onlinecite{Stock06:73,Stock08:77}) at zero field and has been confirmed through the use of constant energy scans at a series of energies.

The low-energy spectrum is qualitatively and quantitatively very similar to that observed in YBCO$_{6.35}$ with its higher superconducting transition temperature of 18 K.  The zero field spin fluctuations display a suppression of scattering at low energies and a broad inelastic peak $\sim$ 2 meV.  A central peak is also observed centered at the elastic line and has been shown previously to display longer correlation lengths than the more heavily doped YBCO$_{6.35}$.~\cite{Yamani07:460}  

On application of a 6 T magnetic field along the $c$ axis, we observe an enhancement of the spin fluctuations at an energy transfer of $\sim$ 0.6 meV (panel $b$).  The relative change in the magnetic intensity is $I(H)/I(0)-1$ $\sim$ 0.1 which is comparable to the changes observed both in the elastic and inelastic channels in the monolayer systems discussed above.  The energy scale of $\sim$ 0.6 meV matches the expected Zeeman splitting for a Cu$^{2+}$ free spin of 0.66 meV.  The relative change and energy scale in YBCO$_{6.33}$ (T$_{c}$=8 K) is similar to that measured in YBCO$_{6.35}$ (T$_{c}$=18 K) indicating that the energy is almost independent of doping.  It also accords with our estimate that the magnetization along the $c$ axis in both experiments is very similar.  Also, as in the YBCO$_{6.35}$ experiment, the field induced resonant excitation has a finite lifetime since it is broader than the resolution bar in Fig. \ref{YBCO_633_energy}.

\section{Discussion and Conclusions}

We have investigated the elastic and inelastic response of the Cu$^{2+}$ spins in superconducting YBCO$_{6.35}$ (T$_{c}$=18 K) and YBCO$_{6.33}$ (T$_{c}$=8 K) to a magnetic field both within the vortex and normal states.  In both phases, we find an enhancement of spectral weight at energy transfers comparable with the Zeeman energy of $\sim$ 0.66 meV.  The relative change in the magnetic intensity is $I(H)/I(0)-1$ $\sim$ 0.1 in both experiments conducted within the vortex and normal phases.  We do not observe a magnetic field response of the elastic central peak neither in the normal nor vortex states.  

In terms of interpreting the inelastic enhancement, it is interesting to review Ref. \onlinecite{Tranquada04:69} in which 10 T affects the inelastic scattering from nearly optimally doped La$_{0.88}$Sr$_{0.12}$CuO$_{4}$. Tranquada \textit{et al.} report a filling in of scattering below an energy transfer of 9 meV and interpret this as evidence for an incommensurate resonance at 9 meV.  A similar experiment described in Ref. \onlinecite{Chang07:98} on La$_{1.895}$Sr$_{0.105}$CuO$_{4}$ reports a filling in of spectral weight at low energies on application of a magnetic field.  This result was taken as evidence for the presence of a renormalized spin-gap.  It is difficult to apply the same reasoning to the field-induced filling in of the spectrum of YBCO$_{6.35}$ (T$_{c}$=18 K) and YBCO$_{6.33}$ (T$_{c}$=8 K) since we observe the same effect with the same energy scale and magnitude in intensity both within the vortex and normal state for a given magnetization along [001].  This similar behavior within vortex and normal states implies the enhancement is not directly driven by superconductivity. It also suggests that the broad peak at $\sim$ 2 meV in heavily underdoped YBCO$_{6+x}$ is not a resonance peak in analogy to heavily doped YBCO$_{6+x}$.  Rather, we think of it as the excitation spectrum of a frozen glass.

Confirming this point, it is not possible to associate the enhanced low-energy fluctuations with the S=1 character theoretically predicted to describe the resonance peak.  In an applied magnetic field, an S=1 state should split into three peaks separated by the Zeeman energy of $\sim$ 0.11 meV/T.  This effect is most readily obseved in quantum dimer systems such as  PHCC (Ref. \onlinecite{Stone07:9}), CuGeO$_{3}$ (Ref. \onlinecite{Enderle01:87}), and (Tl,K)CuCl$_{3}$ (Ref. \onlinecite{Cavadini02:65}).  In underdoped YBCO$_{6+x}$, we do not observe such a triplet splitting but rather an enhancement at a particular energy.  The energy scale of the enhanced scattering matches the expected Zeeman energy both in the vortex and normal states.  Since the field only penetrates along $c$, the field on the spins should be the same for YBCO$_{6.35}$ (T$_{c}$=18 K) at 11 T in the $a-b$ plane and YBCO$_{6.33}$ (T$_{c}$=8 K) at 6 T along the $c$ direction.  The presence of a single excitation which scales with the Zeeman energy implies that the spins causing this excitation have effectively no molecular field and are free spins.  Such a situation could arise from spins located near a broken exchange path resulting from the presence of a hole rich region.  However, since the momentum dependence is the same as the zero field case, the spins must be weakly coupled in a manner similar to the undoped case.  The enhanced fluctuations respond as if the spins were free in terms of energy, yet bound in momentum.  

We now discuss the lack of any field effect at the elastic energy.  Given that the field effect in the monolayer cuprates has been interpreted in terms of a competition between superconducting and antiferromagnetic order parameters, it is most relevant to consider experiments conducted on superconducting and metallic samples.  No strong field effect on the elastic correlations has been reported in underdoped superconducting La$_{1-x}$Ba$_{x}$CuO$_{4}$ (LBCO) with x=0.095 (Ref. \onlinecite{Dunsiger08:77}) while a distinct magnetic field response has been observed in LSCO, La$_{2}$CuO$_{4+y}$, and in more heavily doped LBCO with x=0.12~\cite{Wen08:78}.  It is interesting that all reported magnetic field effects in the monolayer cuprates have been in samples where weak static correlations have been observed in the heavily doped regime (typically $p$ $\sim$ 0.1).  It may be that in the case of heavily underdoped YBCO and also in LBCO, the local antiferromagnetic order is very strong as demonstrated by the significantly long correlation lengths in the $a-b$ plane and the substantial ordered moment at the elastic energy as described in Ref. \onlinecite{Stock08:77}.  The energy scale associated with the static spins (described by the central peak) can be expected to be similar to the exchange constant $J \sim 100$ meV and is much greater than the energy scale of the applied field.  Therefore, the magnetic field (with an energy scale of $\sim$ 0.6 meV) is expected to have little effect on the statically ordered regions of spins characterized by the central peak. 

The issue of microscopic coexistence or phase separation may also be important in the monolayer cuprates.  As noted in Ref. \onlinecite{Dunsiger08:77}, all magnetic field effects have been observed in samples where local probes (such as $\mu$SR) find a significantly reduced magnetic volume fraction. It was suggested in Ref. \onlinecite{Khaykovich02:66} that the field enhanced spin correlations occur primarily in the non magnetic regions where superconductivity dominates.  For heavily underdoped YBCO$_{6+x}$, Ref.\onlinecite{Sanna04:93} has measured a full volume fraction (as in LBCO~\cite{Savici02:66}) and therefore heavily underdoped YBCO$_{6+x}$ is not expected to show a strong magnetic field effect as does LSCO.

In underdoped YBCO$_{6+x}$, the static frozen spins appear to be rigid with no evidence for any spin-flopping with the moment direction orienting perpendicular to the applied field.  If such a situation existed, we would expect an increase in the intensity at $\vec{Q}$=(1/2,1/2,2) of about 10 $\%$, easily observable in our experiment.  Therefore, the static spins are rigidly fixed by internal fields.  This strong static glassy internal field may provide an explanation for the apparent decoupling (ie. no anomaly at T$_{c}$) of the central peak from superconductivity in the heavily underdoped regime.

Our magnetic field results may be understood in terms of a magnetic ground state composed of regions of antiferromagnetic clusters surrounded by hole (or charge) rich regions.  The elastic central peak is characterized by the locally ordered glassy spins in the cluster and we speculate that it is the weakly coupled spins located near the edge of the cluster that give rise to the resonant enhancement of the spin excitations in an applied field.   We have previously suggested a similar model (Refs. \onlinecite{Stock06:73,Stock08:77}) where local antiferromagnetic regions are formed (within the correlation range) separated by metallic regions. This model connects with the one-dimensional stripe structure postulated for LSCO and the nickelates.~\cite{Wakimoto00:61,Woo05:72,Dunsiger07:xx,Berg07:99,Kao05:72}  This result also connects with $\mu$SR studies on YBCO and LSCO that have found evidence for microscopic inhomogeneous superconductivity and multiple transitions or characteristic temperatures.~\cite{Sonier08:101,Sanna04:93}  This model is different than a trivial phase separation of insulating and metallic regions.  If a simple phase separation model were to be invoked, we would expect two distinct spectra - one mimicking the antiferromagnetic insulator with long-ranged antiferromagnetic correlations and one representative of the superconductor. We note that the inelastic response in underdoped YBCO$_{6+x}$ is continuous with energy and temperature, with conservation of spin, and a simple phase separation model of antiferromagnetic and superconducting regions is inconsistent with the available data.

It is worth connecting the ideas here of weakly coupled spins to the behavior of one-dimensional systems with impurity induced edge states.  Similar filling in of gap states has been observed in the Haldane system Y$_{2}$BaNi$_{1-x}$Mg$_{x}$O$_{5}$ as well as in the singlet ground state system SrCu$_{2-x}$Mg$_{x}$(BO$_{3}$)$_{2}$.  The phenomena is associated with edge states (or states near broken chains) introduced through the chemical dopants.~\cite{Kenzelmann03:90,Hara06:97}  Disorder related effects have also been observed in Mg doped CuGeO$_{3}$ where static spin correlations are enhanced as a result of the introduction of weakly bound states from chemical doping.~\cite{Stock05:74}  All of these experiments associated the magnetic field induced effects with weakly coupled spins introduced through chemical disorder which break singlet ground states.  We suggest that the magnetic field enhanced spin fluctuations arise from similar physics to the case of doped one dimensional systems.

In conclusion, we observe a magnetic field induced enhancement of the low-energy spin fluctuations in heavily underdoped YBCO$_{6+x}$.  Through a comparison of results obtained in the normal and vortex states, we conclude that the enhancement is the result of magnetization of weakly coupled spins and not directly related to the suppression of the superconducting order parameter.

\textit{Endnote:} While this paper was submitted to Physical Review, a paper appeared on the arXiv (cond-mat:0902.3335 by Haug \textit{et. al.}) reporting field effects in YBa$_{2}$Cu$_{3}$O$_{6.45}$ with a larger T$_{c}$ = 35 K. Elastic field enhancement occurs unlike the elastic field independence we report for very underdoped YBCO$_{6+x}$.  There is no evidence for Zeeman enhancement of free spins, but only a broad and subtle spectral suppression at larger energies.

\begin{acknowledgements}

We thank R.A. Cowley, R.J. Birgeneau, C. Broholm, G. Xu, Z. Yamani, S. Dunsiger, and J. Tranquada for helpful discussions.  We acknowledge financial support from the Natural Science and Engineering Research Council (NSERC) of Canada and the US National Science Foundation through DMR-0306940.   This work utilized facilities supported in part by the National Science Foundation under Agreement No. DMR-0454672.

\end{acknowledgements}


\end{document}